\documentclass[aps,preprint]{revtex4}

\begin{document}

\title{Observation of spontaneous self-channelling of light in air below the collapse threshold.}
\author{C. Ruiz, J. San Rom\'an, C. M\'endez, V. D\'i az, L. Plaja, I. Arias, L. Roso}
 \affiliation{Departamento de F\'i sica Aplicada, Universidad de
Salamanca, E-37008 Salamanca, Spain}
\date{\today}

\begin{abstract}
We report the observation of the self-guided propagation of 120 fs, 0.56 mJ infrared radiation in air for distances greater than one meter. In contrast to the known case of filamentation, in the present experiment the laser power is  lower than the collapse threshold. Therefore  the counter balance between Kerr self focussing and ionization induced defocussing as the stabilizing mechanism is ruled out. Instead, we find evidences of  a process in which the transversal beam distribution reshapes into a form similar to a Townes soliton. We include numerical support for this conclusion. 
\end{abstract}

\maketitle

Spontaneous guiding of intense femtosecond laser beams is a relevant consequence of propagation in non-linear media. It was first observed by Braun {\em et al } \cite{braun95A} in the form of light filaments extending over a distance of 20 meters. This triggered an intense experimental and theoretical research to better understand the underlying mechanism as well as to develop practical applications  of the phenomena. Fundamentally, the light channels appear in situations in which a compromise between collapse and expansion of the transversal beam profile is possible. In the most known case, self-channeling at high laser powers, these opposite trends come from the self-focussing effect in Kerr media and the dispersion induced by the inhomogeneous refractive index resulting from ionization \cite{dubie04A}. In this case, a filament of light can be spontaneously formed when the self focussing is strong enough to produce the collapse of the beam. The increased intensity in the focus produces ionization and, thus, a change in the refraction index that counterbalances the Kerr focussing. However, this mechanism is not unique:   In this paper we report the observation of this sort of structures for intensities below the critical power of collapse. In this case, ionization does not play a relevant role, and the counter acting mechanism combines the focussing power of the Kerr effect with the beam diffraction. Neglecting the retardation effects of the dispersive medium, the propagation of the laser beam can be described by a non-linear Schr\"odinger equation in the transverse coordinates. In particular, for the cylindric symmetric case we have
\begin{eqnarray}
\label{eq:nls}
i {{\partial} \over {\partial z}} U(r,z,t) & = & - {1 \over {2 k_0}} \left [ {{\partial^2} \over {\partial r^2}}   + {1 \over r} {{\partial} \over {\partial r}} \right ] U(r,z,t) - \nonumber \\
& &  k_0 {n_2 \over n_0} \left | U(r,z,t) \right |^2 U(r,z,t)
 \end{eqnarray}
where $k_0$ is the wave vector of the laser radiation, $n_0$ is the refraction index in air, and $n_2$ is the non-linear refraction index ($3.2 \times 10^{-19}$ cm$^2$/W). Note that Eq. (\ref{eq:nls}) is derived according to the slowly varying envelope  approximation (SVEA). As a result, the temporal coordinate factorizes and every time slice of the electromagnetic pulse evolves independently. $U(r,z,t)$ is defined as the time dependent field envelope $U_0 g(r,z,t)$, so that $U_0^2=E_{in}/V$ with $V=\int rdr dz dt |g(r,z,t)|^2$, $E_{in}$ being the input energy of the beam. A singular aspect of Eq. (\ref{eq:nls}) is the possibility of localized self-trapped solutions, or solitons. In particular, the so-called Townes soliton \cite{chiao64A}. This kind of solutions reflect situations in which diffraction is counterbalanced by self focussing and, therefore, represent transversal self trapping of a portion of the laser beam. The observation of such structure generated after the propagation of an infrared beam through $30 cm$ of BK7 glass has already been reported  in \cite{moll_03A}. However, the stability of this structure was not investigated. This is an important point, since  the Townes solution is known to be unstable under small energy fluctuations of the trapped field \cite{monte04A}. The instability leads to the destruction of the self-trapping in the form of a catastrophic collapse  or the transversal spread of beam. Under these circumstances, the extension of the light channel in experiments depends strongly on the degree of approximation to the actual soliton shape.  On one side, in this paper we report for the first time the observation of the Townes profile in air. Having in mind its unstable nature, the formation of this structure in medium with random fluctuations is a relevant finding. On the other side, we report the stability of the self-trapped channel over more than 1.5 m.

Figure \ref{fig:exp-setup} shows schematically the experimental setup employed in this experiment. A 120 fs, 790 nm  Ti:Sa laser pulse (1 cm transversal FWHM) is focussed in air with a  2.2 m focal lens, right after passing through  pinhole (radius 0.25 cm). A BK7 plate, located at a variable distance form the focal spot, is used to intercept the beam. The transversal energy distribution of the beam is recorded by imaging the plate with a CCD camera. The appropriate control of the energy of the input beam is obtained with the combination of a variable angle $\lambda/2$ plate followed by a linear polarizer with fixed axis. The experiments where performed at 0.11 mJ and 0.56 mJ input energies (measured right after the pinhole). In both cases the associated powers fall below the critical power for collapse due to Kerr self-focussing in air. This has been checked experimentally and also theoretically observing the propagation of the incident beam unmasked by the pinhole (see Fig. \ref{fig:th-intensities}c for the theoretical result). The absence of any radiation channel or filament in this case rules out the possibility of collapse. 

Our experimental results for the beam propagation after the focus are condensed in Fig. \ref{fig:exp-intensities}. Both plots outline the main features of the beam cross section found at different distances from the focussing lens. In particular, the radii corresponding to the widths at $25$, $50$ and $75 \%$ of the maximum energy detected by the CCD camera at different locations of the intercepting plate. The insets correspond to the readout from the CCD camera at a particular position ($\simeq$ 460 cm).   Figure \ref{fig:exp-intensities}a depicts the low energy case (0.11 mJ).   The transversal energy distribution is found to be very close to the expected using the Fresnel diffraction formula, i.e. a ring structure that appear roughly 1 m after the focal spot and can be traced over 3 m before a new diffraction maximum appears at the center of the beam. In contrast, Figure \ref{fig:exp-intensities}b corresponds to the higher energy case (0.56 mJ), and demonstrates completely different features: a pronounced maximum at the center of the beam followed by a more slow decay of the energy for the larger radii. This center structure is very similar to the reported in  \cite{moll_03A}, for a different propagating medium. It  has been identified with a Townes soliton profile immersed in a background which corresponds to the linear diffraction of the lower energy part of the pulse. A relevant aspect described  in Fig. \ref{fig:exp-intensities}b, and the central result of the present paper, is the stability of this distribution over more than 1.5 m.  The stability of the beam channel in our results suggest that the central peak in the energy distribution is rather close to the actual Townes soliton profile. This result corresponds to a self-guiding situation in which an appreciable fraction of the beam energy propagates in a central channel with radius of some hundreds of microns (a situation similar to the filamentation in air with higher intensities). 

Further confirmation of the above interpretation can be drawn form the numerical integration of Eq. (\ref{eq:nls}) with the initial conditions appropriate to describe the pinhole-lens system. As noted above, this equation describes the propagation of a time slice of the electromagnetic pulse. We have chosen, therefore, to investigate the dynamics of the slice corresponding to the maximum intensity. Figures \ref{fig:th-intensities}a and b show the computed energy distributions for the same cases as in figure \ref{fig:exp-intensities}. In the lower energy case, the central part of the beam after the focal point is rapidly depleted, forming the Fresnel diffraction ring. In correspondence with the experimental finding, the higher energy case (Fig. \ref{fig:th-intensities}b) shows a central maximum (marked with an arrow) which decays after one or two meters. While the experimental and theoretical results agree quantitatively in the case of lowest energy, the agreement for 0.56 mJ is only qualitative.  To retrieve the CCD readout form the theoretical point of view, one has to integrate the computations for different time slides accordingly to the actual temporal shape of the pulse. Unfortunately, in our case we have access only to the pulse autocorrelation information, which gives indirect information about the pulse shape. Our checks assuming an ideal gaussian temporal profile do not provide a quantitative agreement satisfactory enough. 

In \cite{moll_03A}, the identification of the Townes profile is obtained by best fit of the structure central maximum with the actual profile of the soliton. Our numerical calculations permit us to be slightly more rigorous. The Townes soliton corresponds to a localized eigenstate of Eq. (\ref{eq:nls}): $U_t(r,z)=\sqrt{I_t(r)} \exp(-i p_t z )$. If we express $U(r,z)=\sqrt{I(r,z)} \exp [i \phi(r,z) ]$ as the central peak of the  solution of Eq. (\ref{eq:nls}) plotted in Fig. \ref{fig:th-intensities}b, the assimilation with the soliton requires: $I(r,z) \simeq I_t(r)$ and $ \partial \phi(r,z) / \partial z \simeq p_t$ (constant). The first condition implies $I(r,z)$ localized and to follow a selfsimilar evolution in $z$  \cite{moll_03A}. The localization of the central maximum in  Fig. \ref{fig:th-intensities}b is evident, and the selfsimilarity along the $z$ coordinate has been found in good approximation. The second condition, i.e. the eigenstate nature, can be considered as the fundamental test. Figure  \ref{fig:th-dfase}a shows  $ \partial \phi(r,z) / \partial z$ for the non-linear case of Fig. \ref{fig:th-intensities}b. The arrow indicates the position of the maximum of the central energy distribution. The flatness of the surface indicates that the condition $ \partial \phi(r,z) / \partial z \simeq p_t$ is well attained in the region where the central structure is defined.

As commented before, the stability of the soliton solution reflects a very subtile balance between diffraction dispersion and Kerr self focussing. Hence, the final achievement of a selftrapped solution depends fundamentally in the initial conditions (amplitude and phase distribution) of the field. In our case, the diffraction pattern generated by the pinhole seems to be of fundamental importance. To show that this is indeed the case, we have performed experiments and simulations of the case corresponding to the 0.56 mJ, but removing the pinhole. Note, that as the field is not screened, the initial beam profile is now gaussian with an energy sensibly higher. In this case the beam expansion after the focal point does not lead to a channel structure, and has the same characteristics as the theoretical plotted in Fig. \ref{fig:th-intensities}b. On the other hand, the phase distribution does not permit the identification of a single eigenstate, as becomes apparent in figure  \ref{fig:th-dfase}b. 

We have reported the spontaneous generation of a narrow channel of radiation in the propagation of a short electromagnetic pulse in air, for energies below the energy threshold in which ionized filaments are possible. We have identified this channel (which extends over more that 1.5 m) as a structure that approximates closely to a
a Townes soliton, hence particularly stable. 

\acknowledgments
This work has been partially supported by the Spanish
Ministerio de Ciencia y Tecnolog\'\i a (FEDER funds, grant
BFM2002-00033)
and by the Junta de Castilla y Le\'on (grant SA107/03). We are grateful to {\em VisionLab} Salamanca for its spirit of collaboration.

\newpage

\begin{figure}
\caption{
Experimetal setup: A Ti:Sa laser beam (790 nm, 120 fs) propagates through a pinhole ( 5 mm diameter) and is focussed by a lens ($f=$2.2 m). The pulse energy is selected by the combination of a $\lambda/2$ plate and a linear polarizer at adequate angles.  After propagating through the focal point, the transversal beam profile is imaged using a BK7 plate and a CCD camera.  Once the energy is measured, the powermeter is removed from the path of the laser beam towards the BK7 plate.}
\label{fig:exp-setup}
\end{figure}

\begin{figure}
\caption{Experimental beam profiles at different distances from the focussing lens for beam energies of 0.11 (a) and 0.56 mJ (b), measured after the pinhole. The data shows the radii corresponding to $25 \%$ (circles), $50 \%$ (squares) and $75 \%$ (crosses) of the intensity maxima at every location.  The insets show the transversal beam profile at the distance marked by the dashed line. }
\label{fig:exp-intensities}
\end{figure}

\begin{figure}
\caption{Transversal field distributions computed from Eq. (\ref{eq:nls}), for the same cases (a) and (b) as in Fig. \ref{fig:exp-intensities}. The black arrow marks the location of the Townes soliton. Plot (c) corresponds to the propagation of the beam in (b)  without interposing the pinhole.}
\label{fig:th-intensities}
\end{figure}

\begin{figure}
\caption{Distribution of the derivative of the field phase along the propagation direction, $ \partial \phi(r,z) / \partial z $, for the non-linear cases with pinhole (figure \ref{fig:th-intensities}b) (a) and without pinhole (figure \ref{fig:th-intensities}c) (b).}
\label{fig:th-dfase}
\end{figure}

\end{document}